\documentclass[conference]{IEEEtran}

\ifCLASSINFOpdf
\else
\fi
\usepackage{amsmath}
\usepackage{soul}
\usepackage{listings}
\usepackage{xurl}
\usepackage[utf8]{inputenc}
\usepackage[T1]{fontenc}
\usepackage{graphicx}
\usepackage{float}
\usepackage{svg}
\usepackage{epstopdf}

\graphicspath{ {./images/} }

\svgsetup{convertdpi=600}

\definecolor{verde}{rgb}{0.25,0.5,0.35}
\definecolor{jpurple}{rgb}{0.5,0,0.35}
\definecolor{darkgreen}{rgb}{0.0, 0.2, 0.13}
\definecolor{lightgray}{rgb}{0.95, 0.95, 0.95}
\definecolor{darkgray}{rgb}{0.4, 0.4, 0.4}
\definecolor{editorGray}{rgb}{0.95, 0.95, 0.95}
\definecolor{editorOcher}{rgb}{1, 0.5, 0} 
\definecolor{editorGreen}{rgb}{0, 0.5, 0} 
\definecolor{orange}{rgb}{1,0.45,0.13}		
\definecolor{olive}{rgb}{0.17,0.59,0.20}
\definecolor{brown}{rgb}{0.69,0.31,0.31}
\definecolor{purple}{rgb}{0.38,0.18,0.81}
\definecolor{lightblue}{rgb}{0.1,0.57,0.7}
\definecolor{lightred}{rgb}{1,0.4,0.5}

\begin{document}
%
\title{Application of HL7 FHIR in a Microservice Architecture for Patient Navigation on Registration and Appointments}
%
%
%



\author{\IEEEauthorblockN{Giovani Nícolas Bettoni}
 \IEEEauthorblockA{\textit{School of Technology}, \\ Pontifical Catholic University of Rio Grande do Sul, \\ Bachelor of Biomedical Informatics, \\ Porto Alegre, Brazil}
\IEEEauthorblockA{Email: $giovani.bettoni$@edu.pucrs.br}

\and

\IEEEauthorblockN{Thafarel Camargo Lobo\\ and Cecília Dias Flores}
\IEEEauthorblockA{Health Information Technology and Management\\  Graduation Program, Federal University of \\ Health Sciences of Porto Alegre, Porto Alegre, Brazil, 90050-170}
\IEEEauthorblockA{Email: $\left \{thafarel, dflores\right \}$@ufcspa.edu.br}

\and

\IEEEauthorblockN{Bruno Gomes Tavares dos Santos}
\IEEEauthorblockA{Psychiatry and Behavioral Sciences Graduation Program, \\ Federal University of Rio Grande do Sul, \\ Bachelor of Biomedical Informatics,\\ Porto Alegre, Brazil}
\IEEEauthorblockA{Email: $gomes.tavares$@ufrgs.br}

\and

\IEEEauthorblockN{Filipe Santana da Silva}
\IEEEauthorblockA{Exact and Applied Sciences Department,\\ Federal University of Health Sciences of Porto Alegre,\\  Porto Alegre, Brazil, 90050-170 \\}
\IEEEauthorblockA{Email: $filipe$@ufcspa.edu.br}

}

\maketitle

\begin{abstract}
Electronic Health Record Systems (EHR-S) are commonly developed in monolithic architectures. This architectural style presents greater complexity and demands more effort when we think of interoperability. A solution proposal is the creation of Microservices that use HL7 FHIR as an interoperability strategy. In this sense, it is presented the development of a prototype, based on a microservices architecture, to act in a real scenario of Patient Navigation (PN). The problem was subdivided into 3 steps: definition of architecture, development and construction of an interface to simulate the role of the navigator. The Patient and Appointment microservices are capable of synchronous communication to query and record information. In general, the implemented architectural style not only isolates information domains but can receive data from multiple sources while maintaining essential functionality. This type of approach plays a crucial role in a hospital environment, specifically in PN, highlighting the importance of the standard and expanding the possibilities for further research to be conducted.
\end{abstract}


%
\IEEEpeerreviewmaketitle

\section{Introduction}
Interoperability in Electronic Health Record Systems (EHR - S) is becoming a functional software requirement and not just an option. We say that different systems interoperate when they can communicate over a common interface \cite{MartinezCosta2011}.

To enable interoperability among EHR-S, it is necessary to incorporate standards such as openEHR \cite{openEHR}, ISO 13606 \cite{ISO2008a}, Health Level 7 (HL7) \cite{HealthLevelSeven2017}, among others. In addition, there are issues such as scalability and maintainability \cite{Hassan2019} that should be addressed.

EHR-S are commonly developed as monoliths \cite{DeLaCruz2011, Oh2015, Hameed2016}. In this approach, there are limitations when allowing data exchange between actors involved, ranging from services isolation to issues regarding data persistence mechanisms. Other issues related to monoliths are their complexity and difficulty to scale \cite{Villamizar}.

These challenges can be minimized with a versatile architecture, such as by means of microservices. These may enable communication between EHR-S and allow strengthening the continuity of care mechanisms, with adequate use of computer resources. Continuity of care is one of the main justifications for interoperability in EHR-S \cite{Kuziemsky2015}.

\begin{figure}[ht]
	\centering
	\includegraphics[width=1\linewidth]{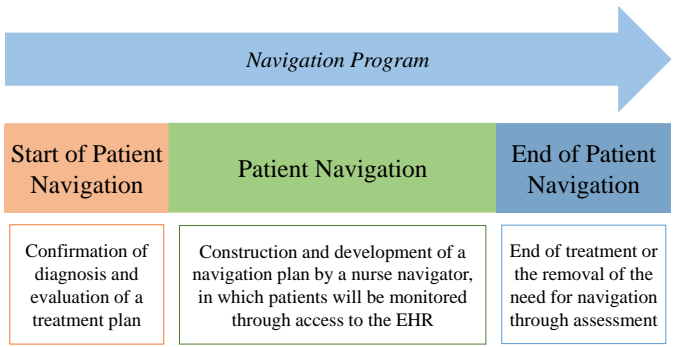}
	\caption{Simplified visualization of process for patient navigation (PN). Adapted from \cite{pautasso2018}.}
	\label{pautasso}
\end{figure}

An example of a solution that requires the establishment of clear mechanisms for continuity of care is Patient Navigation (PN). The PN allows assisting health professionals to provide guidance to patients on the continuous provision of care \cite{pautasso2018}. In general, the PN process consists of 3 main steps (Figure \ref{pautasso}):

\begin{itemize}
    \item Start of Patient Navigation, in which the diagnosis is confirmed and a care plan is evaluated;
    \item Patient Navigation, in which a navigation plan is built by a navigator nurse, where patients will be monitored by means of accessing the EHR data, such as appointment schedules, diagnosis, among others; and
    \item End of Patient Navigation, in which there is the end of treatment, or the removal of the need for navigation through evaluation.
\end{itemize}

Pautasso and collaborators identified requirements that drive the need for the development of a PN application for oncology patients that needed interoperability with existing EHR-S \cite{pautasso2018}. In other words, existing EHR-S could be reused and supplemented with new PN data when necessary. This reuse mechanism can be performed by means of interoperability standards, such as HL7 FHIR \cite{HealthLevelSeven2017}.

Modifying current systems to include new processes, such as PN, could take time and/or resources. To overcome this, new and current EHR-S must incorporate a versatile and connectable strategy to enable interoperability with current or legacy system. In such scenario, it is possible to reuse data from an active or legacy EHR-S. However, this reuse practice could be considerably difficult, as it frequently requires changes to support new care process not yet used in the clinical routine, as in the case of PN.

To enable a solution for such scenario, a viable solution lies in the application of HL7 Fast Healthcare Interoperability Resources (FHIR) \cite{HealthLevelSeven2017}. The HL7 FHIR allows data exchange among different EHR-S in a standardized way. For example, common EHR data like patient registration and scheduling can be exchanged between EHR-S using services in which each of its components are made available as \textit{resources} \cite{HealthLevelSeven2017}. HL7 FHIR is under adoption in diverse clinical scenarios, including current Brazilian National Health Data Network (\textit{RNDS}, Rede Nacional de Dados em Saúde) \cite{Brazil2021}.

Considering HL7 FHIR (in isolation) may help solve some health data communication interoperability issues, we are unable to guarantee the solution of issues that are common to monoliths, as already described. In a recent study, common monoliths issues might be overcome with microservices \cite{Shoumik2017}.

The microservices architectural style is an approach to developing a single system as a set of small services, each running its own process and communicating with lightweight mechanisms, usually an \textit{Application Programming Interface} (API) of HTTP resource. These microservices are:

\begin{itemize}
    \item Built on business rules; and
    \item Independently implemented by fully automated deployment machines.
\end{itemize} 

\noindent There is a minimum of centralized management of these microservices, which can be written in different programming languages and using different data storage technologies \cite{Lewis2014}.

In other words, there is a need for solutions to enable health data interoperability based on the use of HL7 FHIR and microservices. To achieve this scenario, we propose the development of a prototype, based on a microservice architecture that incorporates HL7 FHIR as an interoperability strategy. As a way of describing the viability of the approach, the registration and scheduling of patients will be analyzed, simulating the procedural context of PN.

Our prototype is developed under a microservices architectural approach that uses HL7 FHIR for microservices specification. We took the scheduling and registration of the PN process in order to simulate communication with current HL7 FHIR compliant servers and to demonstrate that the approach is valid even for communication with current HL7 FHIR monolithic servers, as performed by a standard HL7 FHIR implementation from current HAPI SandBox \cite{UniversityHealthNetwork2019}. We envision the combination HL7 FHIR-microservices to promote interoperability and versatility required by EHR-S, current and legacy with proper adaptations (such as data translation routines).

This article is organized as follows. First, we will present the architectural details about which technologies were used for the production and communication through microservices. Next, we will describe how microservices and HL7 FHIR are used to perform patient registration and scheduling. Subsequently, an analysis was carried out relating the results found to those in the literature. Finally, we make some considerations about these findings.

\section{Methods}

As described by \cite{Lobo2020}, there are two communication scenarios, considering data exchange being carried among microservices and an HL7 FHIR server: (i) synchronously or (ii) asynchronously. The incorporation of an (ii) asynchronous approach, which uses the \textit{event-driven} concept \cite{narkhede2017kafka}, has as a first step the consolidation of the (i) synchronous approach for patient registrations and schedules.

\begin{figure}
	\centering
	\includegraphics[width=1\linewidth]{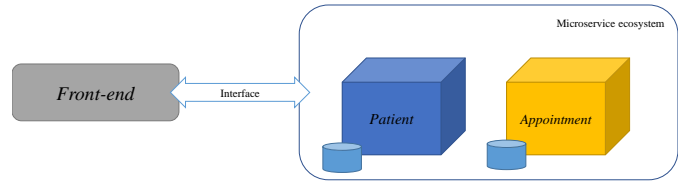}
	\caption{Architecture model.}
	\label{arquitetura}
\end{figure}

This type of approach is conceived from the construction of an architecture that involves specifying the \textit{front-end} and the microservice ecosystem used as part of the \textit{back-end} (Figure \ref{arquitetura}).

In order to develop current approach with HL7 FHIR (R4), we selected all resources related to appointments and scheduling. These are going to be described with more details in Section \ref{resultados}.

\subsection{Development}

\begin{figure}
	\centering
	\includegraphics[width=1\linewidth]{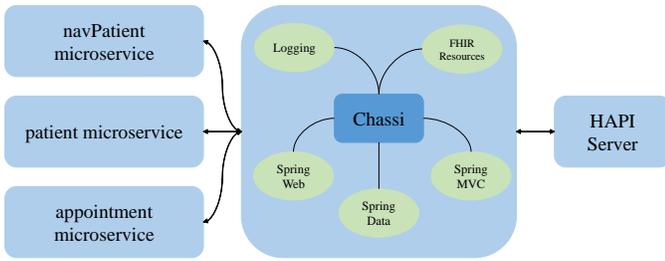}
	\caption{Chassi Pattern. The central box is the chassis, which consists of these subunits in green.}
	\label{chassi}
\end{figure}

Our microsservice development strategy is summarized in Figure \ref{chassi}.

To respect the single responsibility model, we use a Chassis Pattern, composed of five subunits: Logging, Spring MVC, Spring Web, Spring Data and FHIR Resources. The versions of the subunits on the chassis depend on the dependency manager. For the creation of FHIR resources we use version R4.

To facilitate development, we use the Spring Framework \cite{PivotalSoftware2019}. This tool allows the developer to focus directly on the business rules to be developed, without the need to implement connections to databases or web services.

Integration with auxiliary systems happened in 2 ways:

\begin{itemize}
    \item Through REST connections to query data already recorded in the system; and
    \item Access to a NoSQL database \cite{Han2011}.
\end{itemize}

The REST layer was developed with Spring WEB \cite {PivotalSoftware2019}. This tool abstracts the development of translators for information transferred using \textit{JavaScript Object Notation} (JSON), which is used in this work.


The connection via JDBC using Spring Data \cite{PivotalSoftware2019a} was performed with MongoDB \cite{MongoDB2019}. MongoDB records data in JSON format. All data included and used to simulate the PN process were generated manually for demonstration.

\subsection{Communication and HL7 FHIR compliance}
The REST communication incorporates a set of principles that determine how network resources should be defined and addressed to expose microservices functionalities through HTTP operations "consult", "create" or "delete" (GET, POST, PUT or DELETE). These operations allow the exchange of information over the \cite{fielding2000architectural} network.

For this project, we used HL7 FHIR R4 \cite{HealthLevelSeven2017}.
The communication was validated for HL7 FHIR compliance when communication was established with other applications that use the same version of HL7 FHIR. Specifically, the establishment of communication and data exchange was carried out between the presented application and a public reference API that fully implements the HL7 FHIR R4, in our case HAPI \cite{UniversityHealthNetwork2019}. In case of non-compliance, HAPI returns error messages.

\subsection{\textit{Front-end}}

An interface was developed to illustrate how the information generated by the system would be presented to patient navigators.

We chose to use the \textit{Bootstrap} framework - a collection of HTML, CSS and JavaScript tools to create and develop web pages and applications \cite{bootstrap}. Originally created by a designer and developer on Twitter, \textit{Bootstrap} has become one of the most popular front-end frameworks and open-source projects in the world.

Thus, there is the representation of the PN context with the basic data and their processes involved in the registration and scheduling of patients \cite{pautasso2018}. In practice, a \textit{backend} serving the \textit{frontend} (\textit{backend for frontend} or BFF) \cite{Kuziemsky2016, Calcado2015} is responsible for interacting with the user and facilitating data exchange.

To simulate sending and receiving data, as in a real case, the HAPI \textit{Sandbox} \cite{UniversityHealthNetwork2019} was used. It provides a test environment that isolates untested code changes and direct experimentation from the production environment.

\section{Results}
\label{resultados}

Below, we describe the main results found with the development of the proposed approach.

\subsection{HL7 FHIR}

For the development of current strategy, the following HL7 FHIR resources were identified:

\begin{itemize}
    \item \textit{Patient}; and
    \item \textit{Appointment}; and
    \item \textit{Participant}, between \textit{Patient} and \textit{Appointment}.
\end{itemize}

\noindent These resources were included due to the fact that we identified they are enough to represent and exchange data for scheduling and registration. Resources are required for communication and are presented here as well.

\begin{figure}[htp]
	\centering
	\includegraphics[width=0.6\linewidth]{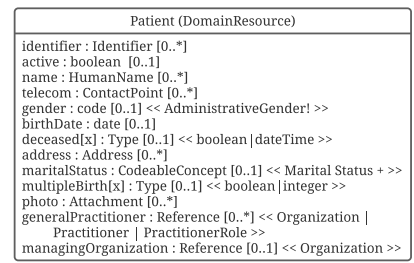}
	\caption{UML description of the resource \textit{Patient}. Source: \cite{HL7Patient}}
	\label{patient}
\end{figure}

The \textit{Patient} resource includes descriptions for the data required to create the patient record, such as the \textbf{identifier}, \textbf{name}, \textbf{date of birth}, \textbf{address}, \textbf{professional} and \textbf{associated organizations}, among others (Figure \ref{patient}).


\begin{figure}[htp]
	\centering
	\includegraphics[width=1\linewidth]{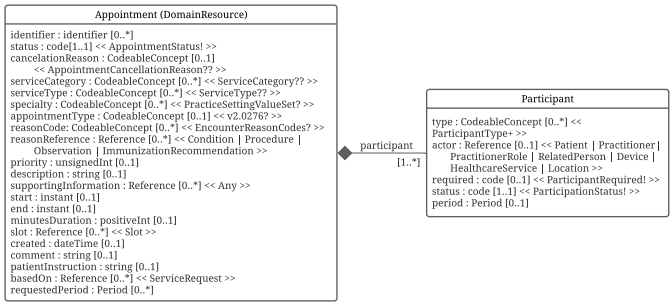}
	\caption{Participation definition for \textit{Patient} in an \textit{Appointment}. Source: \cite{HL7Appointment}}
	\label{appopatient}
\end{figure}

In the \textit{Appointment} resource, data such as \textbf{specialty}, \textbf{start}, \textbf{end}, \textbf{duration}, among others are included. This resource is part of the Workflow component of HL7 FHIR. The relationship between the \textit{Patient} and \textit{Appointment} resources is determined by a participation relationship, which creates a dependency between an individual who will be registered and the respective schedule (Figure \ref{appopatient}).

\begin{figure}[htp]
	\centering
	\includegraphics[width=1\linewidth]{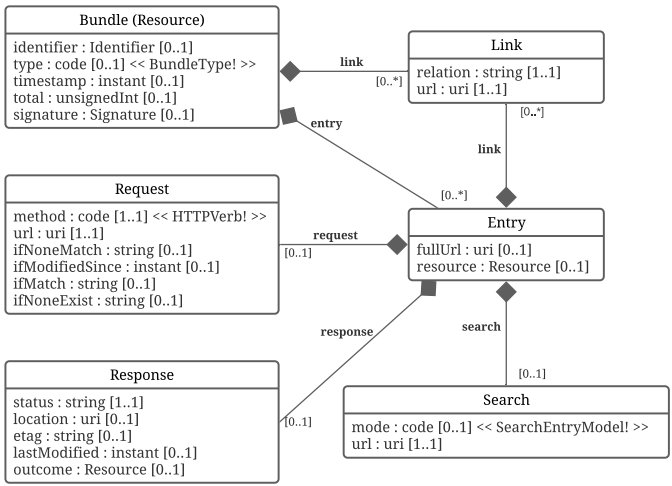}
	\caption{Specification of \textit{Bundle}. Source: \cite{HL7Bundle}}
	\label{bundle}
\end{figure}

Considering that the basis for the implementation of HL7 FHIR incorporates RESTful principles for the communication between a requester and a provider, it is (still) necessary to include certain parts in the REST API. Specifically, it was necessary to include the \textit{Bundle} (Figure \ref{bundle}), which is part of the HL7 FHIR REST API specification. The \textit{Bundle} enables operations such as REQUEST and RESPONSE.

In practice, the \textit{Bundle} provides the specification of the data that must be incorporated in the header of a JSON file, used to establish the communication mechanism.

\subsection{HL7 FHIR and Microservices}

\begin{figure}[htp]
	\centering
	\includegraphics[width=0.5\linewidth]{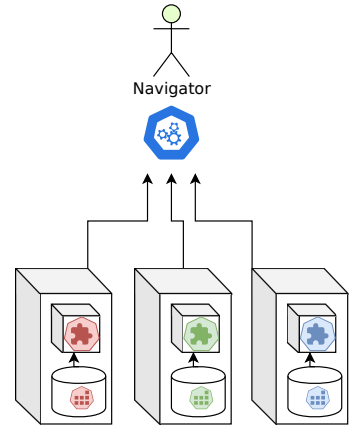}
	\caption{Microservice applications. Source: Elaborated by the authors.}
	\label{nav}
\end{figure}

After defining the resources needed for scheduling and registration, the next step is to adapt the HL7 FHIR to the microservices style. This task lies on defining a persistence strategy to each microservice (Figure \ref{nav}).

\begin{figure}[ht]
	\centering
	\includegraphics[width=1\linewidth]{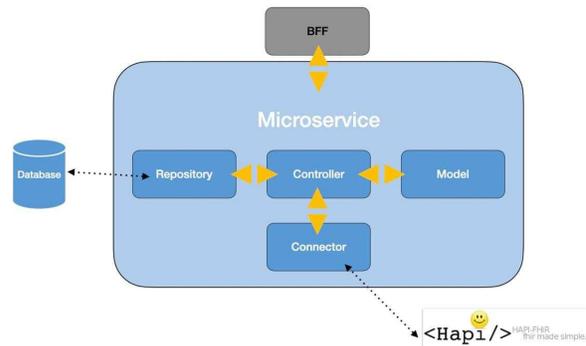}
	\caption{Interpretation by internal of the microservice strategy. Source: Elaborated by the authors.}
	\label{internal}
\end{figure}

We subdivided the microservice into four core components (Figure \ref{internal}):
\begin{itemize}
    \item \textit{Model};
    \item \textit{Controller};
    \item \textit{Connector}; and
    \item \textit{Repository}
\end{itemize}


The \textit{Model} includes the required attributes for the microservice from a given HL7 FHIR resource. For instance, a \textit{Patient} microservice includes attributes such as name, identifier, gender, among others. A \textit{Controller} is typically responsible for preparing a \textit{Model} with data and selecting the front-end but it can also write directly to the response stream and complete the requests. Beyond that, the \textit{Controller} connects with \textit{Repository}, who is responsible for data persistence, and with \textit{Connector}, responsible for the request to a FHIR Sandbox. 

Next, we created the first microservice, named 'Patient'. This one is responsible for making GET requests through the HTTP protocol to a FHIR Sandbox, returning a list of patients. Additionally, it is possible to insert new patients using the POST method, thus composing the \textbf{Patient Record}. 


The second microservice called 'Appointment' has the task of making GET requests through the HTTP protocol to the same FHIR Sandbox, returning a list of schedules and, thus, producing the interaction of \textbf{Scheduling Patients}.

\begin{figure}[htp]
	\centering
	\includegraphics[width=0.7\linewidth]{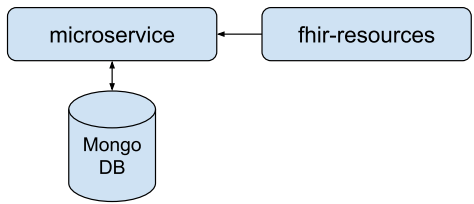}
	\caption{Basic microservice structure created and resources interface. Elaborated by the authors.}
	\label{trem}
\end{figure}

The general structure of the microservice is composed of the microservice itself, along with an interface containing a set of data types that are used for the resource elements, shown in Figure \ref{trem}.

An important detail to consider is the connection establishment mechanism required by HL7 FHIR. To establish a connection, it is necessary to adhere to the message composition protocol. This protocol includes minimal attributes for the specification such as: 

\begin{itemize}
    \item the resource type;
    \item the bundle type (such as document, message, transaction, transaction response, batch, batch response, history list, search results and collection); and
    \item the resource content (such as link, fullUrl, resource, search, request and response).
\end{itemize}


%
%
%
%


\subsection{Prototype: Schedules and consultations for Patients Navigation}

\begin{figure*}[htp]
	\centering
	\includegraphics[width=1\linewidth]{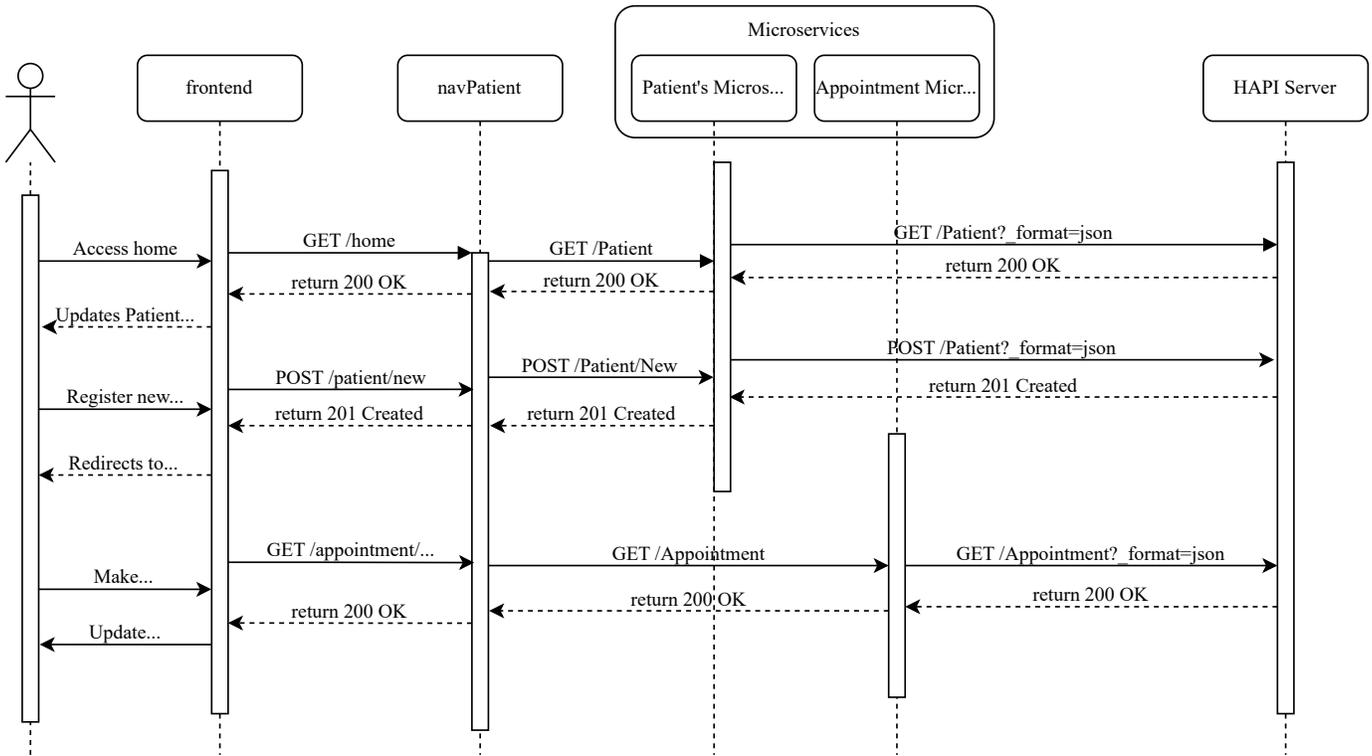}
	\caption{Sequence diagram showing the Communication Process. Source: Elaborated by the authors. Source: Elaborated by the authors.}
	\label{diagrama}
\end{figure*}

The mechanism for registration and scheduling incorporated in this solution is part of the beginning of patient navigation (Figure \ref{pautasso}). The communication established to achieve the interoperability required by the PN mechanism is exemplified in Figure \ref{diagrama}.

\begin{figure}[htp]
	\centering
	\includegraphics[width=1\linewidth]{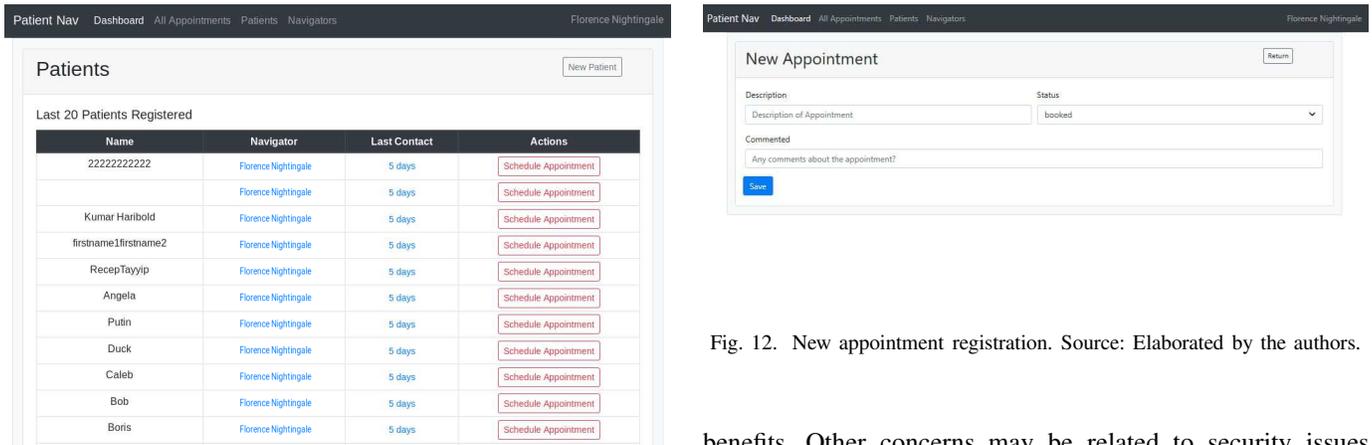}
	\caption{List of patients. Source: Elaborated by the authors.}
	\label{patlist}
\end{figure}

\begin{figure}[htp]
	\centering
	\includegraphics[width=1\linewidth]{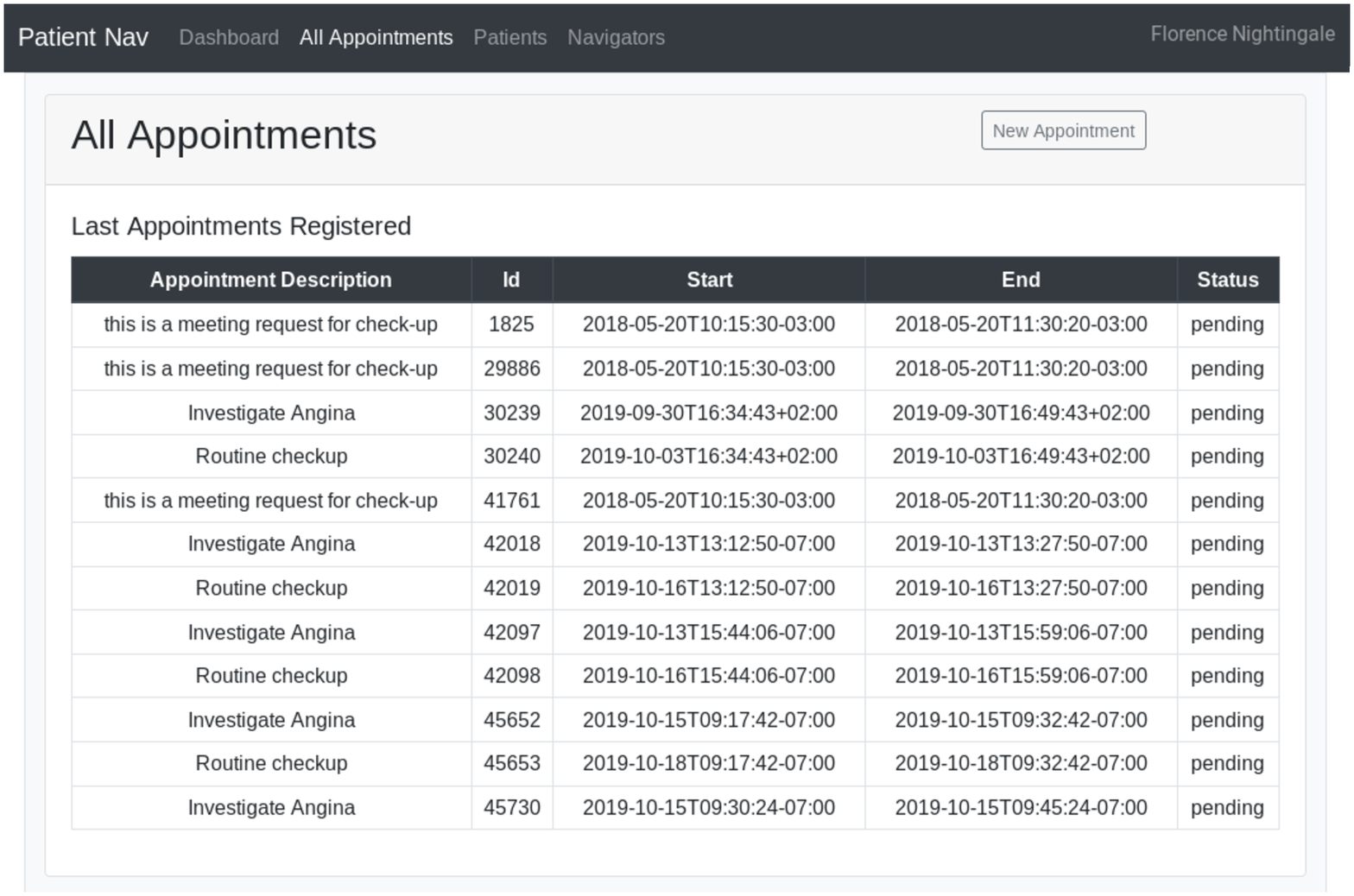}
	\caption{List of appointments. Source: Elaborated by the authors.}
	\label{appo}
\end{figure}

In other words, the navigator will be able to access the \textit{home} screen, which will request the navPatient (BFF) application, through the GET method and this will produce a new request to the \textit{Patient} microservice to retrieve, by means of a HAPI server list, all patients available (Figure \ref{patlist}). This same procedure is replicated or the navigator decides to access the available schedules (Figure \ref{appo}). 

\begin{figure}[ht]
	\centering
	\includegraphics[width=1\linewidth]{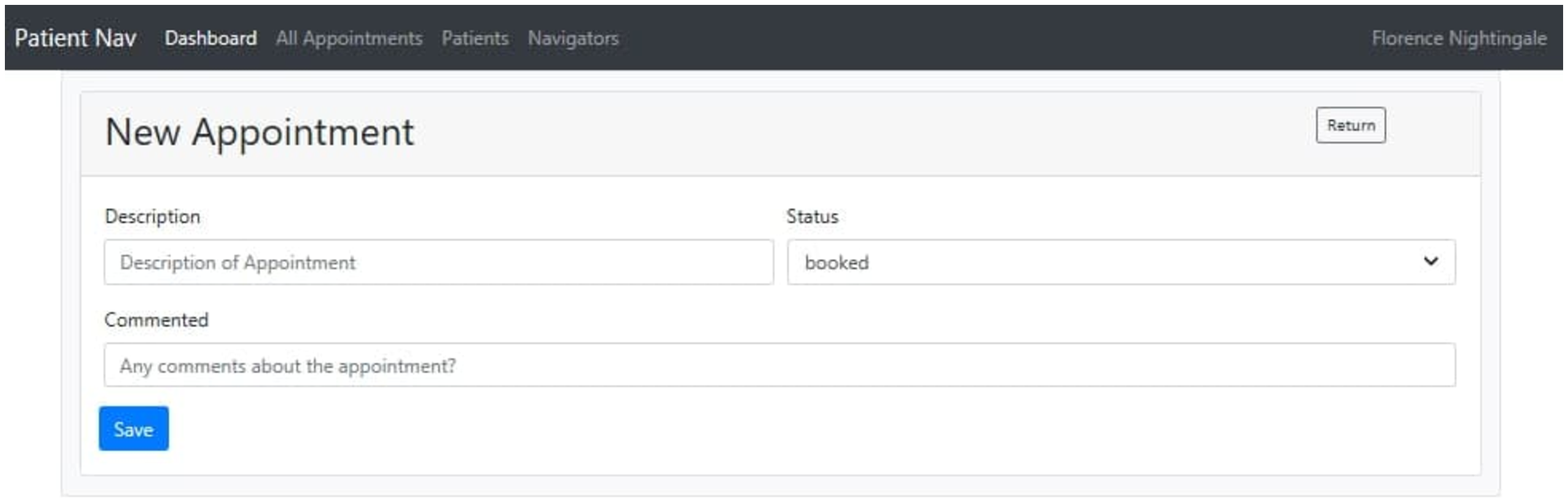}
	\caption{New appointment registration. Source: Elaborated by the authors.}
	\label{newAppo}
\end{figure}

Regarding the scheduling procedure, if an event was already scheduled, the status "booked" is directly retrieved from the server (Figure \ref{appo}). Otherwise, it is necessary create a new appointment (Figure \ref{newAppo}). It is important to note that the appointment user interface is not fully functional at this point. We hope to deliver this interface soon.

\begin{figure}[htp]
	\centering
	\includegraphics[width=1\linewidth]{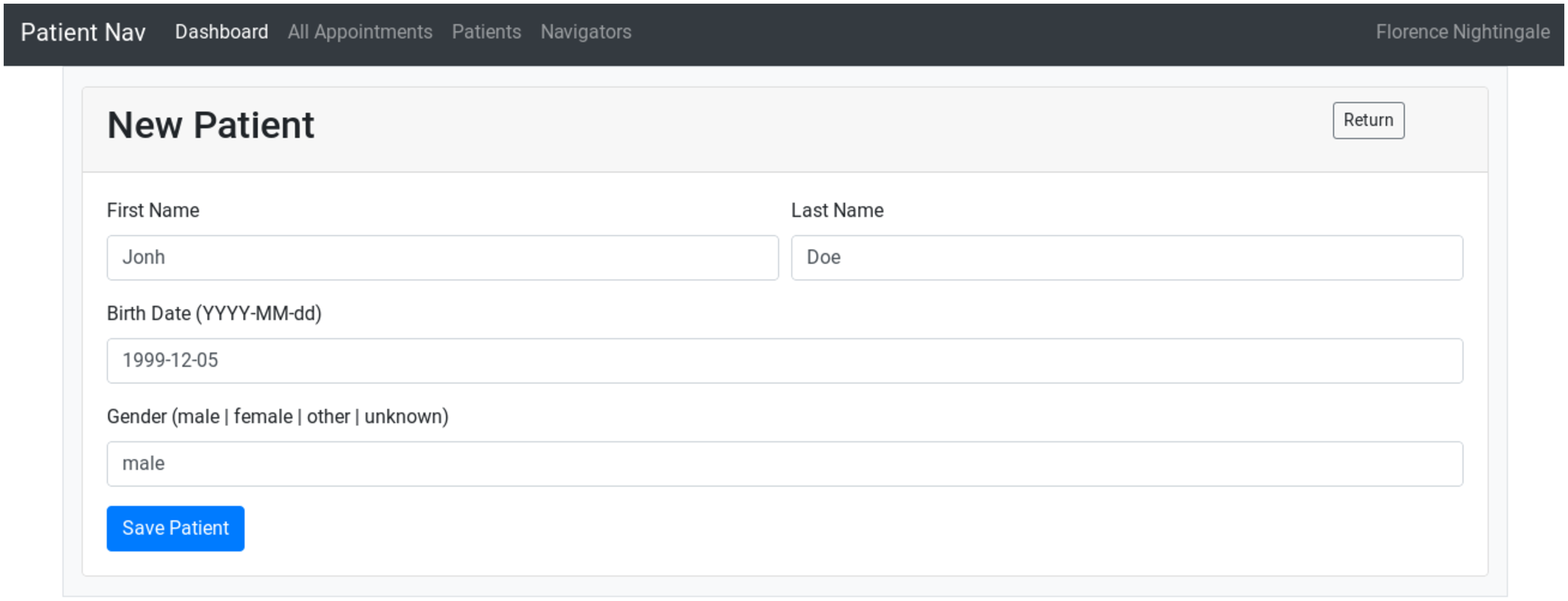}
	\caption{New patient registration. Source: Elaborated by the authors.}
	\label{reg}
\end{figure}

However, to insert a new patient record, the browser will ask the interface to present a form, in which the first name, surname, date of birth and gender must be inserted (Figure \ref{reg}).

Based on this data, a new request will encapsulate the patient object for the \textit{Patient} microservice, which will fill the header with the resource type, and the message body with the data present in the object, when it will be sent to the HAPI server. The confirmation of sending happens through a redirection to the same endpoint; and, the confirmation occurs through the log with the status "201 Created".
\section{Discussion}

According to James Lewis and Martin Fowler \cite{Lewis2014}, the term “microservices” started to be discussed in a workshop. According to these authors, microservices were used to describe an architectural pattern that was already being explored.

In this study, we applied current practices from literature, as in the works of Jamshidi \cite{Jamshidi2018}, Knoche and Hasselbring \cite{Knoche2018} and Newman \cite{Newman:15:MS}. Specifically, with a description of the process of modernizing legacy systems \cite{Knoche2018}. 

We chose to implement synchronous microservices to demonstrate its feasibility for deploying HL7 FHIR based applications; and, as a previous step towards evolving to an event-driven solution, such as presented by \cite{Semenov2019}. However, the synchronous solution may result in downtime \cite{Lewis2014}. This may be related to a high number of calls among microservices. The architecture behind our solution includes both, synchronous and asynchronous, steps and was presented elsewhere \cite{ThafarelCamargoLobo2020}. 

One of the concerns of any project is the ability to evolve. According to the developers, in general, there are two main reasons for the low incorporation of changes \cite{DiFrancesco2019}: 

\begin{itemize}
    \item The deterioration of the internal structure of systems; and
    \item The high number of entry points for client applications.
\end{itemize}

In this type of scenario, the microservice architecture is adequate because it allows a high capacity for evolution, a strong separation of components and a focus on interaction between platforms \cite{Soldani2018}.

On the other hand, and as stated by Jamshidi \cite{Jamshidi2018}, a disadvantage with the increasing popularity of microservices is related to the probability of use in situations where the costs outweigh the benefits. One reason could be that a project would be better developed in a monolithic way. However, this does not mean that the architecture will no longer be modular; only that its modules will not be as isolated as with the use of microservices.

Our work sets out in detail all stages of building microservices with HL7 FHIR, something that is not so accessible in the literature. As Shoumik \cite{Shoumik2017}, we connect each microservice individually and to different ports using a BFF. In \cite{Shoumik2017}, a monolith versus microservice benchmark was carried out to consolidate safety and performance, leaving aside important details about the communication of microservices. In this study, we focus on the application of HL7 FHIR for a real healthcare scenario (process, data model and communication strategy), something that is under investigation by the healthcare software community.


An example of what we find in the community is the proposal of Shaikh \cite{Shaikh2017}. In this work, there is a brief reference that there may be a relationship between microservices and HL7 FHIR. In addition, we find other initiatives that are also attentive to the future of EHR-S architectures \cite{Shoumik2017,Roca2019}, and also propose the use of interoperability standards and microservices. In our work, we removed this idea from the paper and put into practice to explore the viability of this concept for the PN.

Exploring the use of HL7 FHIR, Semenov \cite{Semenov2019} studied a system to support patient's decision. In this work, data aggregation has been proposed to the Patient resource. To perform clinical modeling, a desktop application \cite{Rutten} was used to focus on the functional aspect but hiding the complexity of FHIR. For the problem they were trying to solve, this may be seen as positive, but it is not recommended to initially make abstractions when trying to study the feasibility of migrating the monolith to a microservice architecture \cite{Bucchiarone2018a}. 

The works above reflect a high interest in the theme. However (and as already mentioned), there is little related literature. Existing works do not incorporate details and clear steps. This fact opens up opportunities for further studies to be carried out in this area. In this work we use the REST (synchronous) mechanism for queries and data records encapsulating the HL7 FHIR. In addition, we make HL7 FHIR aspects explicit by declaring resources, the chassis pattern and how they communicate under a synchronous microservices-based architecture. 

However, as in the present study, the application of microservices with EHR-S still needs further evaluation, as described by Bucchiarone \cite{Bucchiarone2018a}. In a real EHR-S context, there is still no evidence of how expensive it is to migrate a monolithic EHR-S to microservices, even with known architectural benefits. Other concerns may be related to security issues regarding data exchange in applications with microservices. 

For instance, \cite{Mateus-Coelho2021} presented some countermeasures or recommendations towards protecting microservice-based systems of attacks, including authentication, authorization, data encryption, single-sign-on solutions, among others. In this study, we did not include any security measures because we were focusing on enabling communication solely. This is the main reason for using manually generated data only to simulate the PN process. However, we are aware of the measures which should be taken for a production environment, considering also aspects from ISO/IEC 27000 standards \cite{ISOIEC2018} (in general) and ISO 27799:2016 \cite{ISO2016} for healthcare.

Even considering the benefits and the incorporation of HL7 FHIR to microservices, some difficulties were encountered. The first (and biggest) of them was the understanding of what a microservice is, how big and how it differs from a service, e.g. when related to a \textit{Service Oriented Architecture} (SOA). 

To assemble the architecture, it took many hours of study to obtain deviations from the monolithic pattern, which is an architectural style under adoption in current applications. Another obstacle to be overcome was the scarcity of practical materials on the subject, as it is still a new topic for the academy, considering this movement begun in the industry \cite{Newman:15:MS}.

Our study faces limitations regarding the evaluation strategy employed. Whereas our goal was just to present the prototype and to see if the approach works, we currently lack results regarding usability and evaluation under a real-world case scenario (i.e. exchanging health data under production environment). Performance evaluation in relation to systems based on Service Oriented Architecture (SOA) based system is out of the scope of the current approach.

\section{Final considerations}
In this work, we present the development of a prototype, based on a microservice architecture that incorporates HL7 FHIR as an interoperability strategy. The feasibility of the approach was determined based on the creation of a solution for registration and scheduling patients as a way to simulate the procedural context of PN.

According to literature, microservices allows some advantages such as:

\begin{itemize}
    \item \textbf{Scalability}: possibility to develop new services without impacting existing ones;
    \item \textbf{Resilience}: a failure is not able to compromise the entire system; and
    \item \textbf{Flexibility}: each microservice can use the language and resources that suits the developers needs.
\end{itemize}

\noindent HL7 FHIR coupled with microservices may deliver an important degree of flexibility to EHR-S development and deploy. HL7 FHIR and microservices might be of interest and require further investigation under the EHR and EHR-S research and industry communities.

From a practical point of view and in accordance with the guidelines identified in literature, the separation of FHIR resources in microservices to compose a PN system is a feasible approach. This is also feasible to our work, which focus in a realistic application.

When compared to the literature, this work fits into a restricted set of few available approaches. Even so, it allows expanding the possibility of conducting new research in which not only the traditional concepts are used (such as REST), but other asynchronous approaches.

As part of an ongoing work \cite{Lobo2020}, there is the intention of incorporating an asynchronous approach to communication and data exchange. One possibility for this, already under analysis and development, is the use of an event-driven approach, by means of a message server, such as Apache Kafka \cite{ApacheSoftwareFoundation2019}. 

Other possible extension is the incorporation of the PN approach in a scenario where the EHR-S server fully incorporates interfaces with the HL7 FHIR to establish direct exchange of EHR data (GET/POST). We envision, in the near future, testing the approach in a production environment in order o evaluate how our architecture behave and supports a proper healthcare setting coupled with a HL7 FHIR production server.

\flushleft{\footnotesize{\textbf{Availability}: \url{https://purl.org/tcc_giovani_bettoni}}}

\bibliographystyle{IEEEtran}

\bibliography{main}

\begin{thebibliography}{10}
\providecommand{\url}[1]{#1}
\csname url@samestyle\endcsname
\providecommand{\newblock}{\relax}
\providecommand{\bibinfo}[2]{#2}
\providecommand{\BIBentrySTDinterwordspacing}{\spaceskip=0pt\relax}
\providecommand{\BIBentryALTinterwordstretchfactor}{4}
\providecommand{\BIBentryALTinterwordspacing}{\spaceskip=\fontdimen2\font plus
\BIBentryALTinterwordstretchfactor\fontdimen3\font minus
  \fontdimen4\font\relax}
\providecommand{\BIBforeignlanguage}[2]{{%
\expandafter\ifx\csname l@#1\endcsname\relax
\typeout{** WARNING: IEEEtran.bst: No hyphenation pattern has been}%
\typeout{** loaded for the language `#1'. Using the pattern for}%
\typeout{** the default language instead.}%
\else
\language=\csname l@#1\endcsname
\fi
#2}}
\providecommand{\BIBdecl}{\relax}
\BIBdecl

\bibitem{MartinezCosta2011}
C.~Martínez-Costa, M.~Menárguez-Tortosa, and J.~T. Fernández-Breis,
  ``{Clinical data interoperability based on archetype transformation},''
  \emph{Journal of Biomedical Informatics}, vol.~44, no.~5, pp. 869--880, oct
  2011.

\bibitem{openEHR}
\BIBentryALTinterwordspacing
S.~Heard and T.~Beale, ``openehr.'' [Online]. Available:
  \url{https://openehr.org/}
\BIBentrySTDinterwordspacing

\bibitem{ISO2008a}
\BIBentryALTinterwordspacing
ISO, ``{ISO 13606-1:2019. Health informatics — Electronic health record
  communication — Part 1: Reference model},'' Geneva, 2019. [Online].
  Available: \url{https://www.iso.org/standard/67868.html}
\BIBentrySTDinterwordspacing

\bibitem{HealthLevelSeven2017}
\BIBentryALTinterwordspacing
H.~L.~S. International, ``Hl7 fhir release 4,'' 2021, last access: Jan 19th,
  2021. [Online]. Available: \url{https://www.hl7.org/fhir/}
\BIBentrySTDinterwordspacing

\bibitem{Hassan2019}
S.~Hassan, R.~Bahsoon, and R.~Kazman, ``{Microservice transition and its
  granularity problem: A systematic mapping study},'' \emph{Software: Practice
  and Experience}, vol.~50, no.~9, pp. 1651--1681, sep 2020.

\bibitem{DeLaCruz2011}
E.~{De La Cruz}, D.~M. Lopez, G.~Uribe, C.~Gonzalez, and B.~Blobel, ``{A
  reference architecture for integrated EHR in Colombia},'' \emph{Studies in
  Health Technology and Informatics}, vol. 169, no.~4, pp. 305--309, 2011.

\bibitem{Oh2015}
S.~Oh, J.~Cha, M.~Ji, H.~Kang, S.~Kim, E.~Heo, J.~S. Han, H.~Kang, H.~Chae,
  H.~Hwang, and S.~Yoo, ``{Architecture design of healthcare
  software-as-a-service platform for cloud-based clinical decision support
  service},'' \emph{Healthcare Informatics Research}, vol.~21, no.~2, pp.
  102--110, 2015.

\bibitem{Hameed2016}
R.~T. Hameed, O.~A. Mohamad, O.~T. Hamid, and N.~Tapus, ``{Design of
  e-Healthcare management system based on cloud and service oriented
  architecture},'' \emph{2015 E-Health and Bioengineering Conference, EHB
  2015}, pp. 1--4, 2016.

\bibitem{Villamizar}
M.~Villamizar, O.~Garces, H.~Castro, M.~Verano, L.~Salamanca, R.~Casallas, and
  S.~Gil, ``{Evaluating the monolithic and the microservice architecture
  pattern to deploy web applications in the cloud},'' in \emph{2015 10th
  Computing Colombian Conference (10CCC)}, M.~Sanchez and O.~Gonzalez,
  Eds.\hskip 1em plus 0.5em minus 0.4em\relax Bogota, Colombia: IEEE, sep 2015,
  pp. 583--590.

\bibitem{Kuziemsky2015}
C.~E. Kuziemsky, ``{A Multi-Tiered Perspective on Healthcare
  Interoperability},'' in \emph{Standards and Standardization},
  M.~Khosrow-Pour, Ed.\hskip 1em plus 0.5em minus 0.4em\relax IGI Global, 2015,
  ch.~56, pp. 1166--1181.

\bibitem{pautasso2018}
F.~Pautasso, ``Navegadores de pacientes: Implantação de um programa de
  navegação para a oncologia,'' Master's thesis, Universidade Federal de
  Ciências da Saúde de Porto Alegre, Porto Alegre, 2018.

\bibitem{Brazil2021}
\BIBentryALTinterwordspacing
Brazil and {Ministry of Health}, ``{RNDS - Rede Nacional de Dados em
  Sa{\'{u}}de},'' 2021. [Online]. Available: \url{https://rnds.saude.gov.br/}
\BIBentrySTDinterwordspacing

\bibitem{Shoumik2017}
F.~S. Shoumik, M.~I. M.~M. Talukder, A.~I. Jami, N.~W. Protik, and M.~M. Hoque,
  ``{Scalable micro-service based approach to FHIR server with golang and
  No-SQL},'' in \emph{20th International Conference of Computer and Information
  Technology, ICCIT 2017}, vol. 2018-January.\hskip 1em plus 0.5em minus
  0.4em\relax Dhaka, Bangladesh: IEEE, dec 2018, pp. 1--5.

\bibitem{Lewis2014}
\BIBentryALTinterwordspacing
J.~Lewis and M.~Fowler, ``{Microservices: a definition of this new
  architectural term},'' 2014. [Online]. Available:
  \url{https://martinfowler.com/articles/microservices.html}
\BIBentrySTDinterwordspacing

\bibitem{UniversityHealthNetwork2019}
\BIBentryALTinterwordspacing
{HAPI FHIR}, ``{HAPI-FHIR},'' 2019. [Online]. Available:
  \url{https://hapifhir.io/}
\BIBentrySTDinterwordspacing

\bibitem{Lobo2020}
\BIBentryALTinterwordspacing
T.~C. Lobo, G.~N. Bettoni, F.~S. da~Silva, R.~C. Caregnato, and C.~D. Flores,
  ``{Enabling communication among EHR systems with microsservices and HL7
  FHIR},'' in \emph{Actas de SABI2020}, Piri{\'{a}}polis, Uruguai, 2020, p.
  247. [Online]. Available: \url{http://sabi2020.com/proceedings-actas/}
\BIBentrySTDinterwordspacing

\bibitem{narkhede2017kafka}
N.~Narkhede, G.~Shapira, and T.~Palino, \emph{Kafka: the definitive guide:
  real-time data and stream processing at scale}.\hskip 1em plus 0.5em minus
  0.4em\relax " O'Reilly Media, Inc.", 2017.

\bibitem{PivotalSoftware2019}
\BIBentryALTinterwordspacing
{Pivotal Software}, ``{Spring Framework},'' 2019. [Online]. Available:
  \url{https://spring.io/}
\BIBentrySTDinterwordspacing

\bibitem{Han2011}
J.~Han, E.~Haihong, G.~Le, and J.~Du, ``{Survey on NoSQL database},''
  \emph{Proceedings - 2011 6th International Conference on Pervasive Computing
  and Applications, ICPCA 2011}, pp. 363--366, 2011.

\bibitem{PivotalSoftware2019a}
\BIBentryALTinterwordspacing
{Pivotal Software}, ``{Spring Data},'' 2019. [Online]. Available:
  \url{https://spring.io/projects/spring-data}
\BIBentrySTDinterwordspacing

\bibitem{MongoDB2019}
\BIBentryALTinterwordspacing
MongoDB, ``{MongoDB},'' 2019. [Online]. Available:
  \url{https://www.mongodb.com/}
\BIBentrySTDinterwordspacing

\bibitem{fielding2000architectural}
R.~T. Fielding and R.~N. Taylor, \emph{Architectural styles and the design of
  network-based software architectures}.\hskip 1em plus 0.5em minus 0.4em\relax
  University of California, Irvine Doctoral dissertation, 2000, vol.~7.

\bibitem{bootstrap}
\BIBentryALTinterwordspacing
P.~H. {Galante, Andres Harborow, Bardi Cuppens, Martijn Otto, Mark Lauke},
  T.~McDonald, and S.~Yoshida, ``{Bootstrap},'' 2011. [Online]. Available:
  \url{https://getbootstrap.com/docs/4.4/about/overview/}
\BIBentrySTDinterwordspacing

\bibitem{Kuziemsky2016}
C.~E. Kuziemsky and L.~Peyton, ``{A framework for understanding process
  interoperability and health information technology},'' \emph{Health Policy
  and Technology}, pp. 1--8, 2016.

\bibitem{Calcado2015}
\BIBentryALTinterwordspacing
P.~Cal{\c{c}}ado, ``{The Back-end for Front-end Pattern (BFF)},'' 2015.
  [Online]. Available:
  \url{https://philcalcado.com/2015/09/18/the_back_end_for_front_end_pattern_bff.html}
\BIBentrySTDinterwordspacing

\bibitem{HL7Patient}
\BIBentryALTinterwordspacing
HL7, ``{Resource Patient}.'' [Online]. Available:
  \url{http://hl7.org/fhir/R4/patient.html}
\BIBentrySTDinterwordspacing

\bibitem{HL7Appointment}
\BIBentryALTinterwordspacing
------, ``{Resource Appointment}.'' [Online]. Available:
  \url{http://hl7.org/fhir/R4/appointment.html}
\BIBentrySTDinterwordspacing

\bibitem{HL7Bundle}
\BIBentryALTinterwordspacing
------, ``{Resource Bundle}.'' [Online]. Available:
  \url{https://www.hl7.org/fhir/bundle.html}
\BIBentrySTDinterwordspacing

\bibitem{Jamshidi2018}
P.~Jamshidi, C.~Pahl, N.~C. Mendonca, J.~Lewis, and S.~Tilkov,
  ``{Microservices: The journey so far and challenges ahead},'' \emph{IEEE
  Software}, vol.~35, no.~3, pp. 24--35, may 2018.

\bibitem{Knoche2018}
H.~Knoche and W.~Hasselbring, ``{Using Microservices for Legacy Software
  Modernization},'' \emph{IEEE Software}, vol.~35, no.~3, pp. 44--49, may 2018.

\bibitem{Newman:15:MS}
S.~Newman, \emph{Building Microservices: Designing Fine-Grained Systems},
  1st~ed.\hskip 1em plus 0.5em minus 0.4em\relax O'Reilly Media, February 2015.

\bibitem{Semenov2019}
I.~Semenov, R.~Osenev, S.~Gerasimov, G.~Kopanitsa, D.~Denisov, and
  Y.~Andreychuk, ``{Experience in Developing an FHIR Medical Data Management
  Platform to Provide Clinical Decision Support},'' \emph{International Journal
  of Environmental Research and Public Health}, vol.~17, no.~1, p.~73, dec
  2019.

\bibitem{ThafarelCamargoLobo2020}
\BIBentryALTinterwordspacing
T.~C. Lobo, G.~N. Bettoni, F.~S. da~Silva, R.~C. Caregnato, and C.~D. Flores,
  ``{Enabling communication among EHR systems with microsservices and HL7
  FHIR},'' in \emph{Actas de SABI2020}, Piri{\'{a}}polis, Uruguai, 2020, p.
  247. [Online]. Available: \url{http://sabi2020.com/proceedings-actas/}
\BIBentrySTDinterwordspacing

\bibitem{DiFrancesco2019}
P.~{Di Francesco}, P.~Lago, and I.~Malavolta, ``{Architecting with
  microservices: A systematic mapping study},'' \emph{Journal of Systems and
  Software}, vol. 150, pp. 77--97, 2019.

\bibitem{Soldani2018}
J.~Soldani, D.~A. Tamburri, and W.~J. {Van Den Heuvel}, ``{The pains and gains
  of microservices: A Systematic grey literature review},'' \emph{Journal of
  Systems and Software}, vol. 146, pp. 215--232, 2018.

\bibitem{Shaikh2017}
F.~A. Shaikh, B.~J. Kolowitz, O.~Awan, H.~J. Aerts, A.~von Reden, S.~Halabi,
  S.~A. Mohiuddin, S.~Malik, R.~B. Shrestha, and C.~Deible, ``{Technical
  Challenges in the Clinical Application of Radiomics},'' \emph{JCO Clinical
  Cancer Informatics}, no.~1, pp. 1--8, 2017.

\bibitem{Roca2019}
S.~Roca, J.~Sancho, J.~Garc{\'{i}}a, and {\'{A}}.~Alesanco, ``{Microservice
  chatbot architecture for chronic patient support},'' \emph{Journal of
  Biomedical Informatics}, p. 103305, oct 2019.

\bibitem{Rutten}
\BIBentryALTinterwordspacing
M.~Rutten, ``{Forge FHIR}.'' [Online]. Available:
  \url{https://fire.ly/products/forge/}
\BIBentrySTDinterwordspacing

\bibitem{Bucchiarone2018a}
A.~Bucchiarone, N.~Dragoni, S.~Dustdar, S.~T. Larsen, and M.~Mazzara, ``{From
  Monolithic to Microservices: An Experience Report from the Banking Domain},''
  \emph{IEEE Software}, vol.~35, no.~3, pp. 50--55, may 2018.

\bibitem{Mateus-Coelho2021}
N.~Mateus-Coelho, M.~Cruz-Cunha, and L.~G. Ferreira, ``{Security in
  Microservices Architectures},'' \emph{Procedia Computer Science}, vol. 181,
  no. 2019, pp. 1225--1236, 2021.

\bibitem{ISOIEC2018}
\BIBentryALTinterwordspacing
ISO/IEC, ``{ISO/IEC 27000:2018 Information technology — Security techniques
  — Information security management systems — Overview and vocabulary},''
  Geneva, pp. 1--34, 2018. [Online]. Available:
  \url{https://www.iso.org/standard/73906.html}
\BIBentrySTDinterwordspacing

\bibitem{ISO2016}
\BIBentryALTinterwordspacing
ISO, ``{ISO 27799:2016. Health informatics — Information security management
  in health using ISO/IEC 27002.}'' Geneva, 2016. [Online]. Available:
  \url{https://www.iso.org/standard/62777.html}
\BIBentrySTDinterwordspacing

\bibitem{ApacheSoftwareFoundation2019}
\BIBentryALTinterwordspacing
{Apache Software Foundation}, ``{Apache Kafka},'' 2019. [Online]. Available:
  \url{https://kafka.apache.org/}
\BIBentrySTDinterwordspacing

\end{thebibliography}

\end{document}